\documentclass[seceq]{ptptex}

\usepackage{graphicx}
\usepackage{wrapft}

\newcommand{\bx}{{\mib x}}
\title{%        %You can use \\ for explicit line-break
Exploration of Multi-dimensional Density of States 
by Multicanonical Monte Carlo algorithm
}

\author{%       %Use \scshape  for the family name
Yukito \textsc{Iba}$^{1,}$\footnote{E-mail: iba@ism.ac.jp}
and Hisanao \textsc{Takahashi}$^{2}$%
}

\inst{%         %Affiliation, neglected when [addenda] or [errata]
$^1$The Institute of Statistical Mathematics, Tokyo 106-8569, Japan \\ 
$^2$267-5 Kita-Akitsu Tokorozawa-shi Saitama-ken 359-0038, Japan}

\abst{%         %this abstract is neglected when [addenda] or [errata]
Multi-dimensional density of states provides a useful 
description of complex frustrated systems.
Recent advances in Monte Carlo methods enable efficient calculation
of the density of states and related quantities, 
which renew the interest in them. 
Here we calculate density of states on
the plane {\sf (energy, magnetization)} for an Ising Model with three-spin
interactions on a random sparse network, 
which is a system of current interest both in physics
of glassy systems and in the theory of error-correcting codes.
Multicanonical Monte Carlo algorithm is successfully applied, and
the shape of densities and its dependence on the 
degree of frustration is revealed. Efficiency of 
multicanonical Monte Carlo is also discussed with the shape
of a projection of the distribution simulated by the algorithm. 
}
\begin{document}

\maketitle

\section{Introduction}

Given a system of the energy $E(\bx)$, the density of states
of physical quantities $A_0(\bx), A_1(\bx), \ldots$ is
defined as
$$
D(A_0=a_0, \, A_1=a_1, \,
\ldots \, ) \, = \, \sum_{\bx} \,\, \bigg \{
\delta(A_0(\bx)-a_0) \times \delta(A_1(\bx)-a_1) \times
\cdots \, \bigg \}_,
$$
where $\delta$ is defined as usual: $\delta(a)=1$ if $a=0$, else
$\delta(a)=0$. Usually, one of the $A_i$s' is 
the energy $E(\bx)$ itself. When we set 
$A_0(\bx)=E(\bx)$ and ignore other $A_i$s',
it gives the original definition of a univariate 
``density of state'' $D(E)$, which is the 
normalization constant of microcanonical distribution 
$
P^{mc}(\bx) = \delta(E(\bx)-E_0)/D(E_0) {}_.
$
Hereafter we are mostly interested in 
bivariate cases where $A_0(\bx)=E(\bx)$,
$A_1(\bx)=M(\bx)$, where $M(\bx)$ is an order parameter
of a system, for example, magnetization of an Ising ferromagnet. 

Recently, an introduction of 
extended Monte Carlo 
methods~\cite{iba}, especially 
multicanonical Monte Carlo
algorithm~\cite{Berg,
2dim,iba}, makes calculation of 
multi-dimensional densities of states a realistic 
choice for the analysis of complex
probability distributions.
There is a special interest in studying the cases with a first-order
transition with latent heat with this approach. 
Existence of the latent heat indicates a rapid drop of the
entropy of the system in ordering and suggests 
the difficulty in searching 
ground states. It will be interesting to 
investigate the nature of ordering process 
through the calculation of the multi-dimensional density 
of states. In this paper we study Ising models
with three-spin interactions on random sparse 
networks (``Sourlas code'')~\cite{KNW}
and calculate density of the states 
and related quantities by multicanonical Monte Carlo.
%%For this model first-order transition, 
%%we successfully calculate density of the states, 
%%multicanonical distribution by using 
%%multicanonical Monte Carlo.

%We also touch on the comparison to 10-states Potts model 
%on the square lattice. 
%%There seem, however, not many
%%studies which utilized the computational
%%advantage of multicanonical algorithm.
%In the fields of statistical physics of 
%spin systems and information processing,
%few works explicitly  discuss multi-dimensional 
%desisity of states. 
%%Sheto {\it et al}~\cite{2-dim}
%%investigate a density on the  plane 
%%{\sf (energy, magnetization)} for an Ising
%%model on square lattice. 
%Some of the studies on lattice protein models, especially 
%%Chikenji~{\it et al.} discuss multi-dimensional 
%%density of states of a lattice protein model 
%%in terms of the structure of folding path (``funnels'').
%Our results mostly consist of ``illustrations'', 
%each of which corresponds to a sigle random realization of a given system 
%size. 
%Although more systematic study is required for any definite conclusion, 
%we hope our preliminary results activate interest in this direction.
%Microcanonical approaches based on densities of states are 
%useful both in physics and statistical information processing.  

\section{Multicanonical Monte Carlo}

Details of multicanonical Monte Carlo algorithm
are found in the references~\cite{Berg,iba}. The essence of the algorithm 
is dynamical Monte Carlo sampling 
with a weight $1/\tilde{D}(E)$, where $\tilde{D}(E)$ is an
approximation of $D(E)$. It causes a nearly uniform
distribution of the energy $E$ within an interval. 
The density $\tilde{D}(E)$ is estimated by repeated 
``preliminary runs'' of the simulation (a simple 
method called ``entropic sampling''~\cite{Lee} is used here).
%Usually, the results with the weight $1/D(E)$ is
%converted to canonical averages by a reweighing procedure.
%In our case, 
The bivariate density $D(E,M)$ is 
obtained by
$\log D(E,M)= \log W(E,M) + \log(\tilde{D}(E)) + const. {}_,$
where $W(E,M)$ is the frequencies of the values
$(E,M)$ appeared in a simulation
with the weight $1/\tilde{D}(E)$. 
We can also reconstruct a conditional density 
$D(M|E)$ by $\log D(M|E)= \log D(E,M) - \sum_M \log D(E,M) {}_.$
$D(E|M)$ is proportional to the projection on 
$(E,M)$ plane of the distribution
sampled by multicanonical Monte Carlo with 
an ideal weight $\tilde{D}(E)=D(E)$.
%which will be useful to study the behavior of 
%multicanonical Monte Carlo.

The advantage of multicanonical Monte Carlo is that
relaxation in small $E$ (and also very high $E$) region is
greatly facilitated compared with canonical or microcanonical 
simulations. An interesting question is 
when multicanonical Monte Carlo
performs better than the other methods with extended ensembles, 
parallel tempering~\cite{iba} and 
%(replica exchange Monte Carlo) and 
simulated tempering~\cite{marinari,iba}. 
It is usually believed that multicanonical Monte Carlo 
performs better in cases with first-order transitions 
with latent heat, because multicanonical ensemble 
can contain states that rarely appear 
in a canonical distribution of any
temperature, and relaxation is speed up 
with a vanishing critical nucleus size.
%The ``gap'' on the
%energy axis in the coexistence region is ``filled'' 
%by the learning of $\tilde{D}(E)$
%in the preliminary runs and relaxation is speed up 
%with a vanishing critical nucleus size.
This picture, however, seems not to be tested
enough. We will discuss it in the later section with
our example. 

\section{Ising Model with three-spin
interactions}

Let us consider an Ising model with three-spin
interactions:
$$
P_\beta(\bx) \propto \exp \left ( \beta \sum_{(ijk) \in G} 
J_{ijk} x_i x_j x_k \right ) {}_.
$$
Here $\sum_{(ijk) \in G}$ denotes the summation on 
the edges of a graph $G$. The energy and the magnetization 
is defined as $E= \sum_{(ijk) \in G} J_{ijk} x_i x_j x_k$, 
and $M= \sum_i x_i$, respectively. Several different
definitions of a random network $G$ is possible. Here we consider
a simplest case to simulate: The probability
of the presence of the edge is independent and takes the same value
for all pairs of the vertices on $G$.
Total number of edges are fixed to a given number $I$, but the number $K$ 
of the edges that contains a vertex is not fixed. Hereafter we set $J_{ijk}= \pm 1$.
%%(It differs fromreferences~\cite{}.).

\begin{wrapfigure}{c}{12cm}
\begin{minipage}{6cm}
\includegraphics[width=5cm]{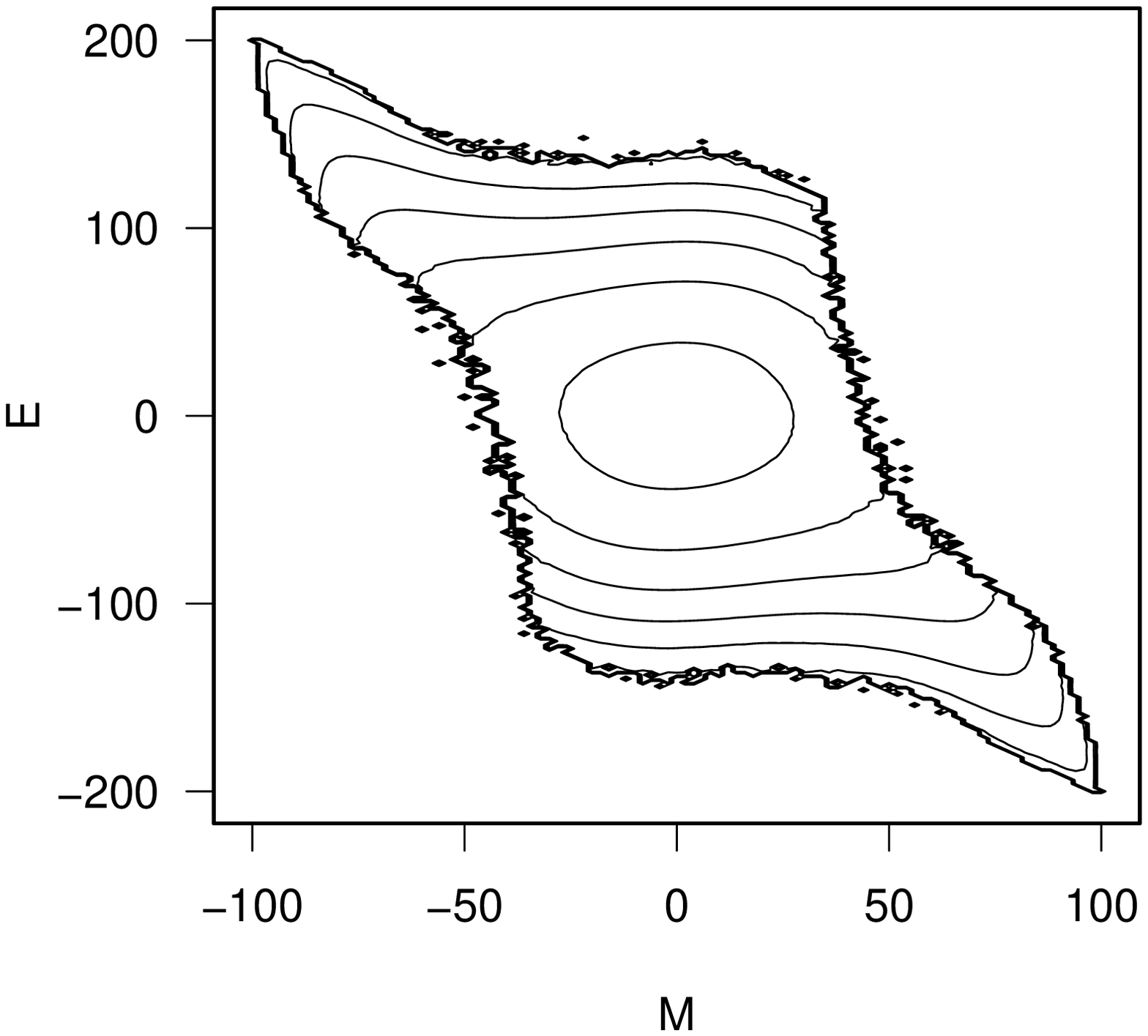}
\caption{Density of States: $\log D(E,M)$} \label{fig_dens1}
\end{minipage}
\begin{minipage}{6cm}
\includegraphics[width=5cm]{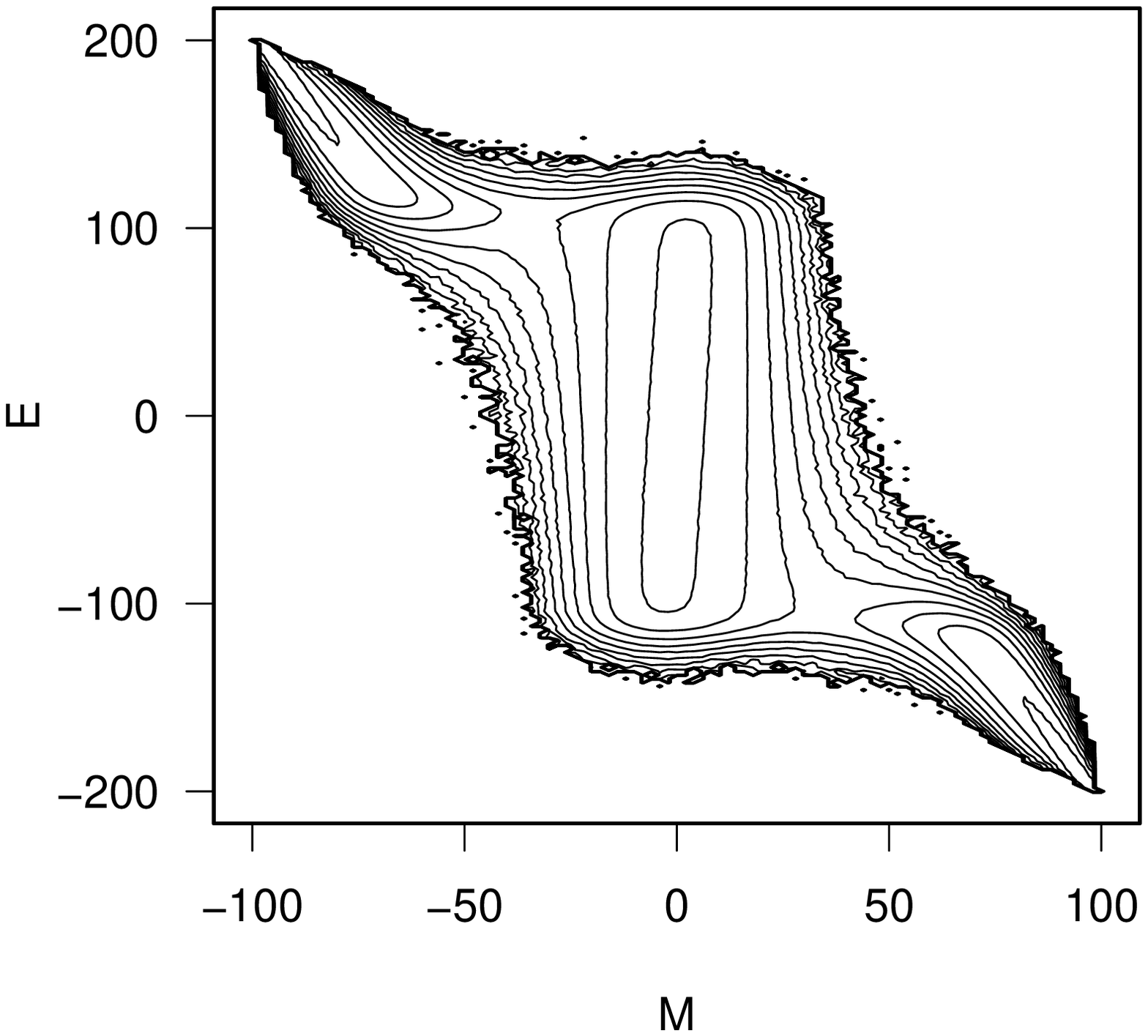}
\caption{Multicanonical Density: $\log D(M|E)$} \label{fig_mul1}
\end{minipage}
\end{wrapfigure}

An example of the results are shown in Figs.~\ref{fig_dens1} 
and \ref{fig_mul1} for a ferromagnetic 
model ($J_{ijk} \equiv 1$) of the size $N=100$ on a random graph $G$ with
$I=200$ edges. Density of states $D(E,M)$ 
and the projection  $D(E|M)$ of multicanonical density are
plotted in Figs.~\ref{fig_dens1}
and \ref{fig_mul1}. The vertical and horizontal axes
corresponds to the energy $E$ and the magnetization $M$. 
In Figs.~\ref{fig_dens1}~and~\ref{fig_mul1}, an interval of 
the contours corresponds to the change of the density by the factors 
$10^4$ and $10^{0.5} \sim 3.2$, respectively. 
%levels=4*c(-10:0) dens
%0.5*c(-40:0) multi
%\begin{center}
%\includegraphics[width=12cm]{./3body/100_200_0/IDEM100_200_0.ps}
%\end{center}
%%\begin{center}
%%\end{center}
In Fig.~\ref{fig_dens1}, states around $(E,M)=\pm (200,100)$ 
at the ends of ``arm''s observed in both sides of the body 
correspond to the ordered states,
while the peak at the center corresponds to the paramagnetic state. 
Point symmetry of the density corresponds to $S_i \rightarrow -S_i$, 
$(E,M) \rightarrow (-E,-M)$ symmetry of the model. 
%A pair of ridges
%is observed along the arms, each of which connects the paramagnetic state to 
%one of the ordered state. 
In Fig.~\ref{fig_mul1}, three peaks of
the density are observed, one of which is the paramagnetic state at the
center, and the others correspond to ordered states. Multiple peaks 
indicate that metastabilty appears even with multicanonical Monte Carlo. 
%although the mixing at lowest energy will be
%improved compared with canonical or microcanonical simulation.

%levels=0.5*c(-80:0),
%oldversion:levels=c(-40:0)
\begin{wrapfigure}{c}{12cm}
\begin{minipage}{6cm}
%\begin{center}
\includegraphics[width=5cm]{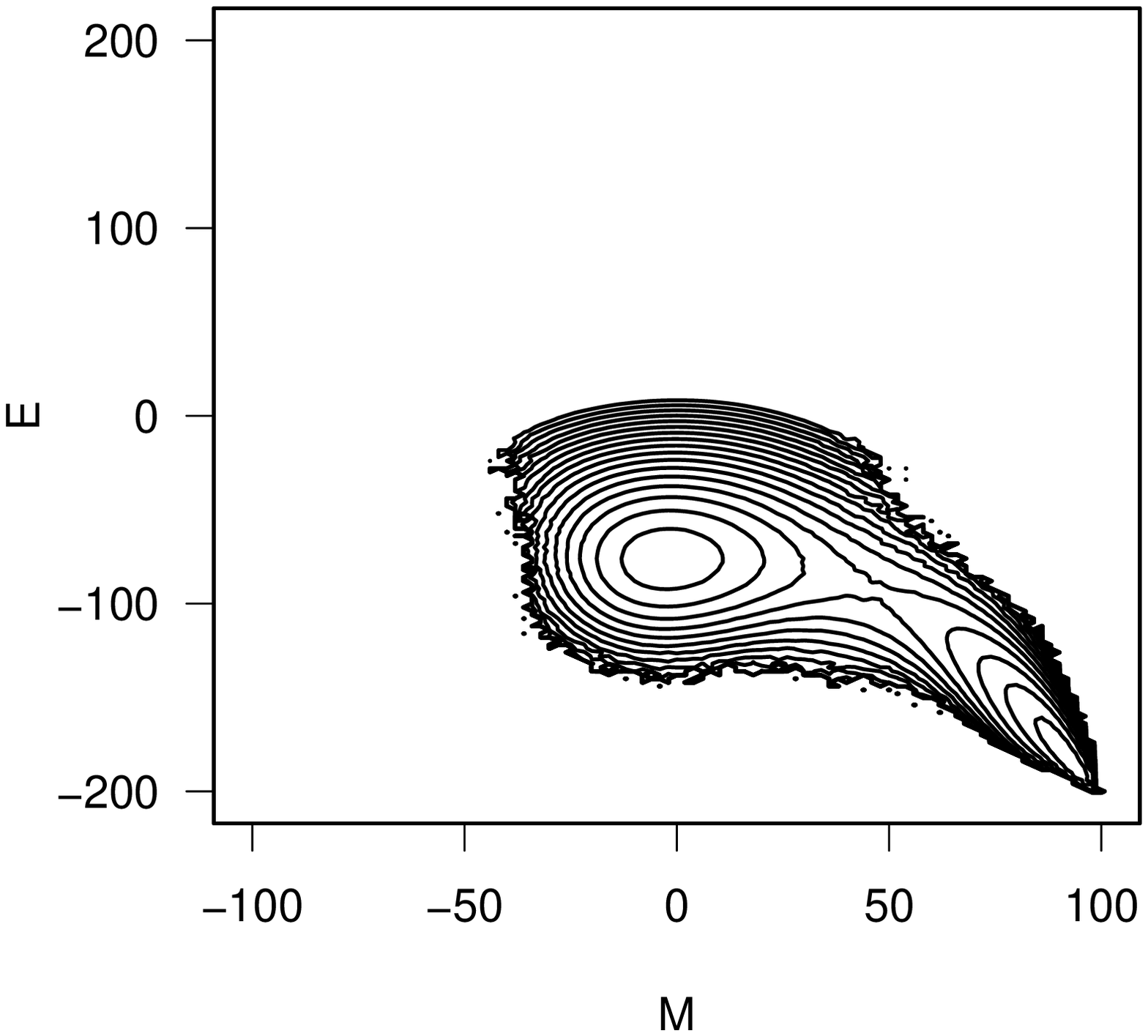}
\caption{Distribution of $(E,M)$ at $\beta=0.4$.} 
\label{fig_gibbs}
%\end{center}
\end{minipage}
\begin{minipage}{6cm}
%\begin{center}
\includegraphics[width=5cm]{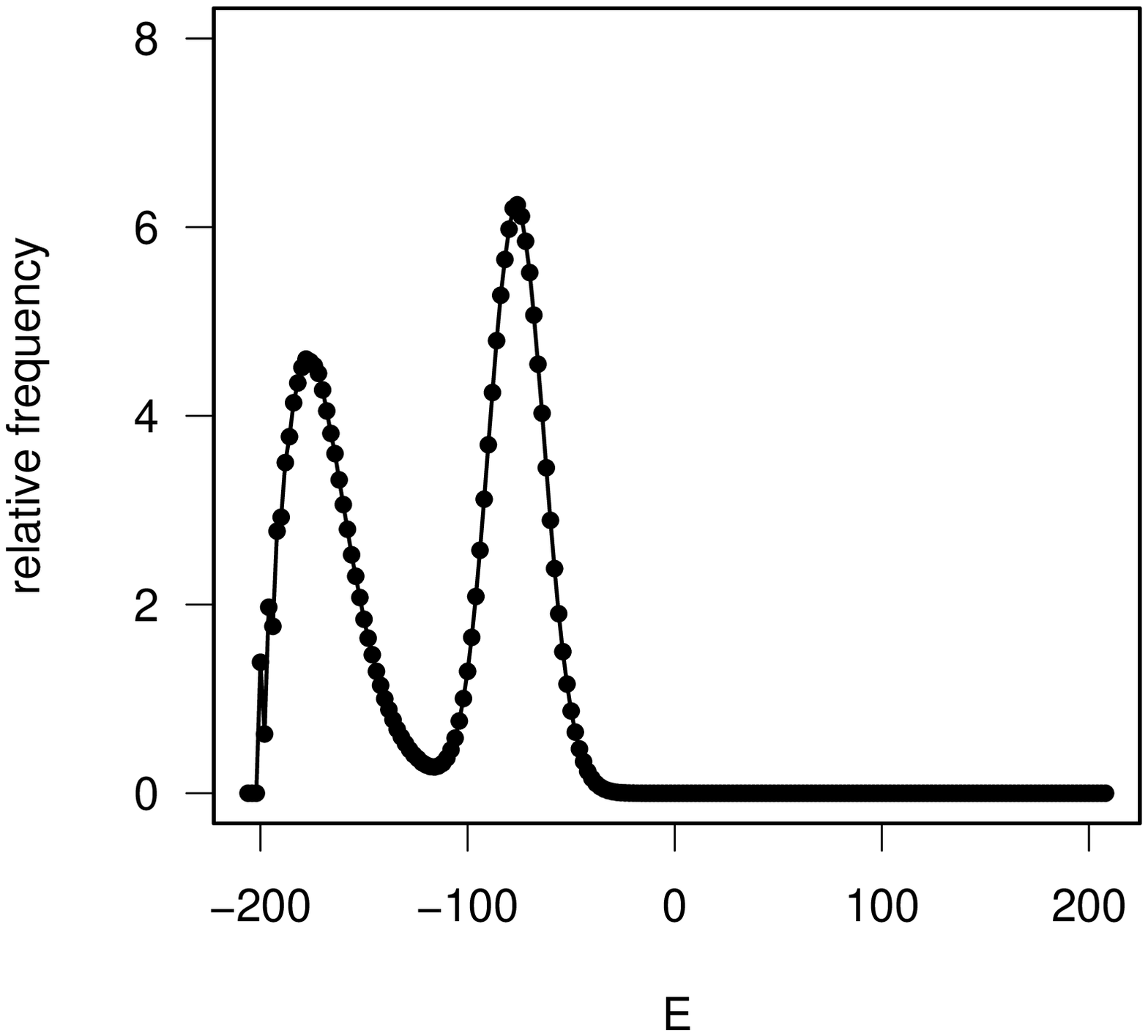}
\caption{Distribution of $E$ at $\beta=0.4$.}
\label{fig_energy}
%\end{center}
\end{minipage}
\end{wrapfigure}

In Figs.~\ref{fig_gibbs}~and~\ref{fig_energy}, projections
of a canonical distribution $P_\beta (x) \,\, (\beta=0.4)$ on 
the $(E,M)$ plane  
and on the $E$ plane are shown, respectively.  
In Fig.~\ref{fig_gibbs}, the contour interval 
and axes are the same as those in Fig.~\ref{fig_mul1}.
Two peaks are again seen in
both figures, which agree with an existence of 
the first-order transition with latent heat~\cite{KNW}.

\begin{wrapfigure}{c}{14cm}
%%\begin{center}
\begin{minipage}{4.5cm}
\begin{center}
\includegraphics[width=4.5cm]{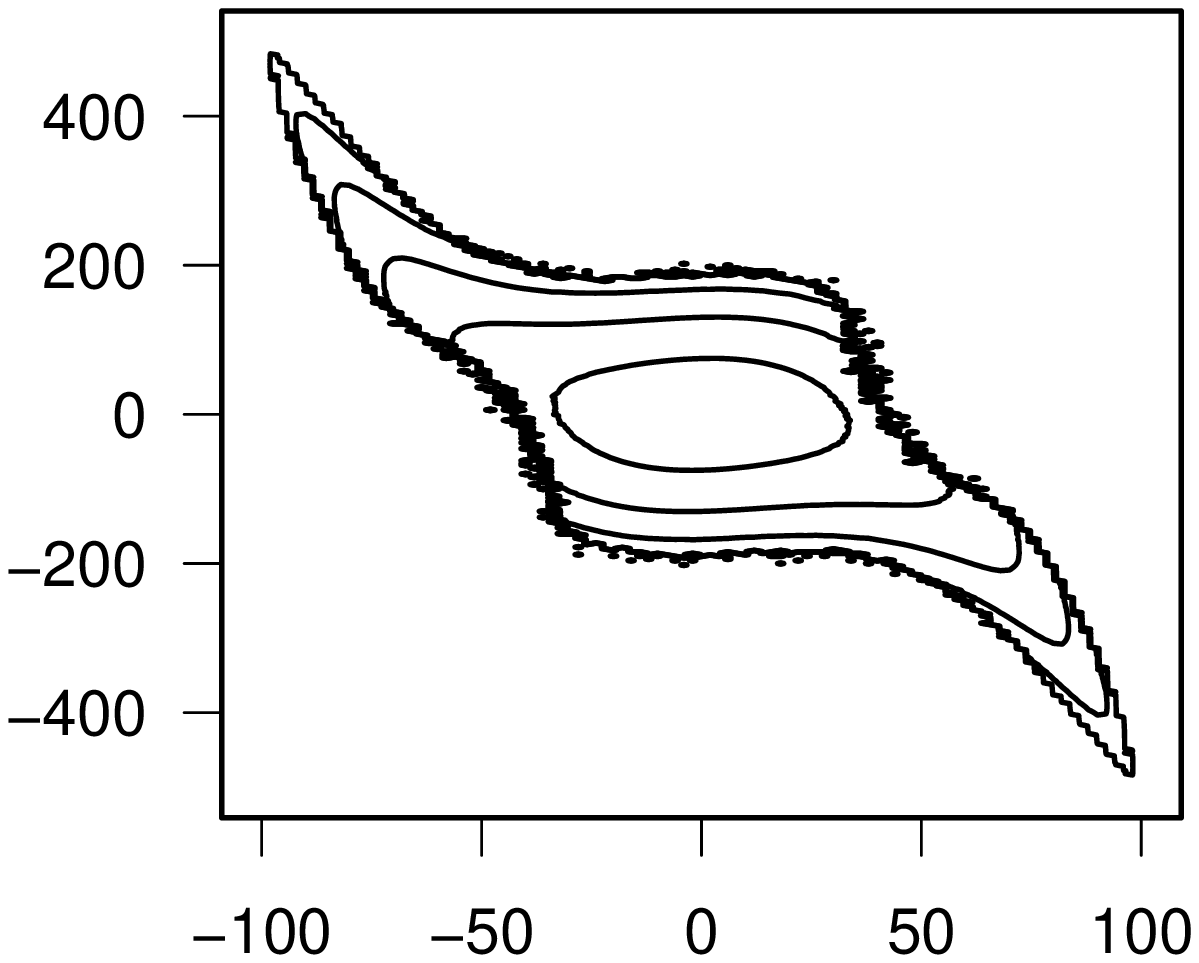}
%\vspace*{-0.8cm}
\caption{$p=0$} \label{fig_100_500_0}
\end{center}
\end{minipage}
\begin{minipage}{4.5cm}
\begin{center}
\includegraphics[width=4.5cm]{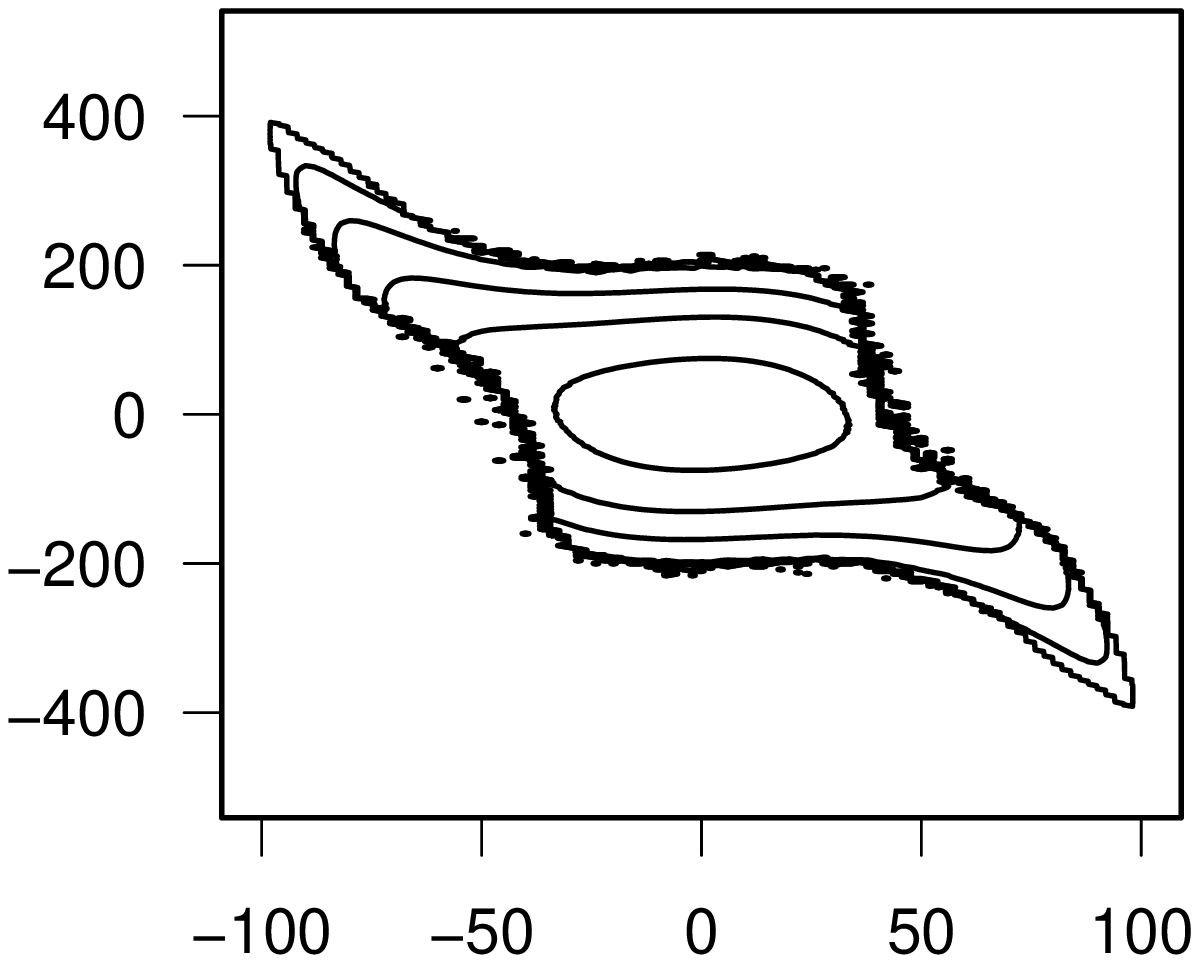}
\vspace*{-0.8cm}
\caption{$p=0.1$} \label{fig_100_500_50}
\end{center}
\end{minipage}
\begin{minipage}{4.5cm}
\begin{center}
\includegraphics[width=4.5cm]{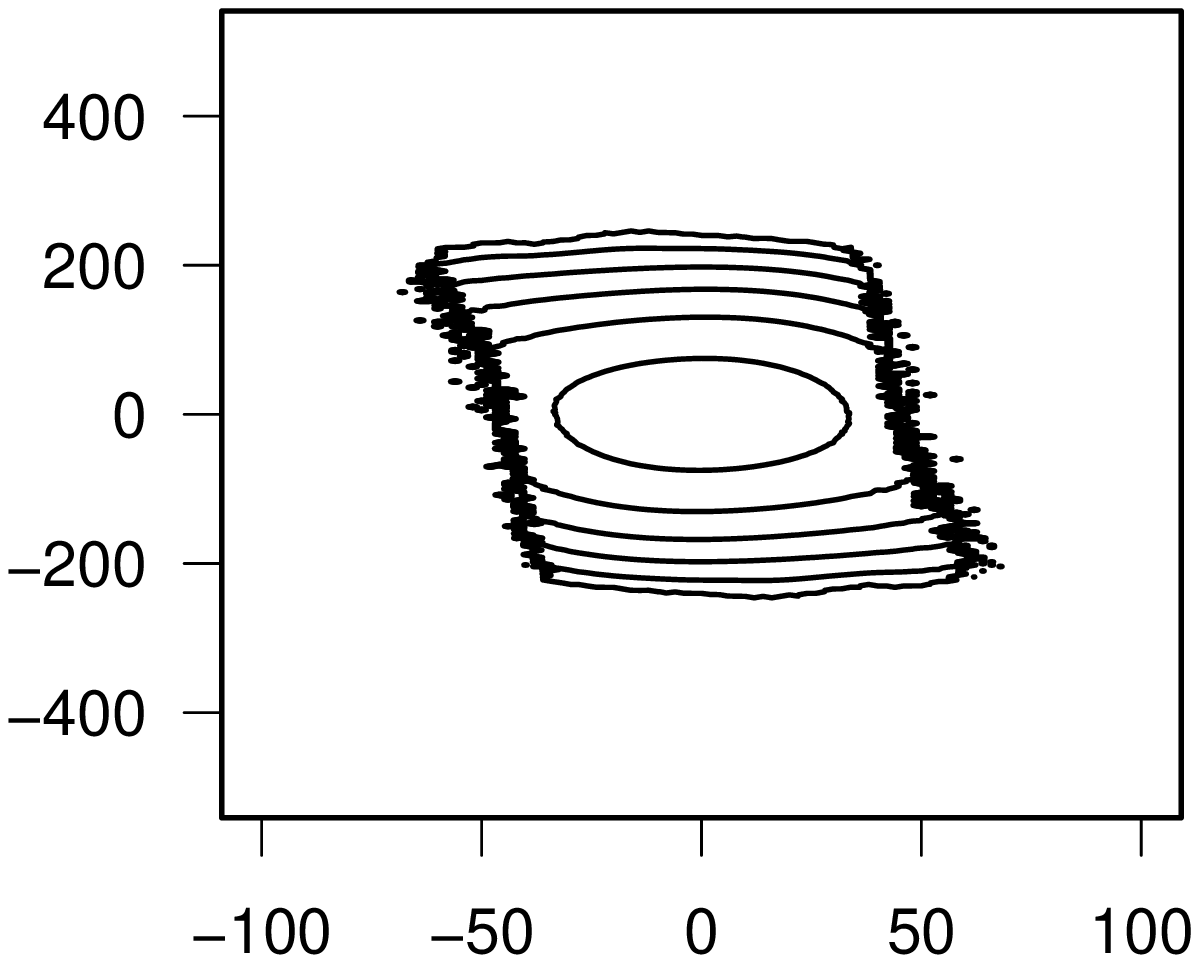}
\vspace*{-0.8cm}
\caption{$p=0.36$} \label{fig_100_500_180}
\end{center}
\end{minipage}
%%\end{center}
\end{wrapfigure}

The changes of the shape of density of states when we add 
frustration are shown in 
Figs.~\ref{fig_100_500_0}-\ref{fig_100_500_180}. Models
with $N=100$ and $I=500$ are investigated for
the ratio $p$ of negative bonds
0, 0.1, and 0.36. The contour interval and axes are the same as those 
in Fig.~\ref{fig_dens1}.
Increasing the concentration of frustration, 
the maximum value of magnetization that appears
with a probability higher than a threshold reduces
and the ``arm''s shrink. At the value $p=0.36$, there are 
no more evident arms, which indicate ordered states 
do not exist at any temperature. 
For each of these examples,
$1.0 \sim 1.5 \times 10^9$ MCS are 
required for the estimation of $\tilde{D}(E)$ and $\sim 10^8$ MCS are used 
for the computation of the physical quantities. Typical CPU times 
are $10 \sim 15$ hours with a 3.2GHz Pentium4 chip.

\section{Discussion and Future Problems}

%The results presented here is limited to results for
%a single sample of fixed size, and careful examination of
%scaling limit and averages over the randomness
%is required for any definite conclusion. 

The result shown in Fig.~\ref{fig_mul1} 
give rise to the possibility that (univariate)
multicanonical Monte Carlo may not have a 
clear advantage over parallel tempering 
in this example. A natural question is that 
how multicanonical density looks like 
for other systems with first-order transition with latent heat.
Our preliminary results on 10-states Potts model 
on the square lattice indicate that multicanonical
density projected on a {\sf (energy, order parameter)} plane
shows smoother behavior compare with those of Ising models
with three-spin interactions, although a saddle point is
still observed. Detailed comparison is, however, left for future studies,
as well as computation of a thermodynamic limit of the densities
and direct comparison between multicanonical and parallel tempering algorithm. 
%Properties of states around $E \sim 120$ 
%in  Fig.~\ref{fig_mul1} also attracts our attention as a future
%problem. These states rarely appear in a canonical ensemble but 
%can be contained in multicanonical or Tsallis ensembles.

Another interesting issue is the use of two-dimensional
multicanonical Monte Carlo~\cite{2dim} 
with the weight $1/\tilde{D}(E,M)$ that
approximates inverse of the bivariate density $D(E,M)$. 
It enables efficient sampling of small $D(E,M)$ regions, and, hopefully,
further enhance the mixing. 
Our experience, however, 
shows that it requires considerably many iterations 
for the estimate of the bivariate $\tilde{D}(E,M)$ by entropic sampling. 
Some improvements in the learning stage should be
introduced in future studies.

%there seems many different
%cases -- In some cases, excessive thermal fluctuations destroy the
%germ of the order and make the effort of finding ground states
%difficult. In other cases, the landscape is golf-course like, i.e., 
%optimization problem is very difficult to solve. 

%\section*{Acknowledgements}
%We would like to thank ...........

%\appendix
%\section{First Appendix} %Empty argument \section{} yields `Appendix'. 
%
%\section{Second Appendix}

\end{document}